\documentclass[aps]{revtex4}
\begin{document}
\title{ Gauge Theory of Gravitation }
\author{Diaa A. Ahmed}
\address{e-mail vasishta@dds.nl}
\date{Rotterdam November 1997}

\begin{abstract}
Gravitation as a quantum topological phenomenon on a large scale. Introducing the action of gravitation into functional space
-through a phase angle- requires a change in our understanding of both gauge theory and general relativity. This displays its unity
with the rest of gauge interactions and demonstrates that the manifold is compact. 
 {04.60.-m} {11.10.-z}
\end{abstract}
\maketitle

\section{Introduction}
The dimensionless coupling $G (x) = k m / r$ , plays the role of phase angle of cosmological origin; coupling between dipole
moments. We can derive its action in functional space by introducing it into the group structure, $exp \{ i g T^{\alpha} .
{\omega}^{\alpha} (x) \}$ . We may use the notation $\{^{\alpha}_{\beta}\} = g^{\alpha\mu} g_{\mu\beta}$ , and its
contracted form $\{  \}$, to represent this phase angles and to facilitate the calculations in terms of basis $\{ \omega^{\alpha}
(p) \}$ , $\{ e_{\beta} (p) \}$ , and phase angles $< {\omega}^{\alpha} | e_{\beta} >$ . \\ 
The phase angles exist in functional space and its contractions equal in the geometric language to its projections on the metric. 
Action of the phase angles reconstruct the wavefront and this determines the geometric structure; of the manifold and the fields. \\ 
We will now consider that these phase angles perform rotations that are subject to same algebraic scheme as the fields $W$ . 
This sort of rotations will mix up fields in order to bring in the transformations on these fields. Furthermore the transformations
will not change the structure of physical law. This means that the phase $\omega^{\alpha} (x)$ , that is involved is a general
one and $G (x)$ , is associated with the same algebraic structure of $T^{\alpha}$. This will permit these transformations to induce
the appropriate transformations on the fields. Therefore we will assume,

\begin{equation}
exp \{ i G (x) T^{\alpha} . {\omega}^{\alpha}_{\beta} (x) \} ,
\end{equation}

\begin{equation}
[ T^{\alpha} , T^{\beta} ] = i f^{\alpha\beta\gamma} T^{\gamma} ,
\end{equation}

The infinitesimal transformations;

\begin{equation}
( I + i T^{\alpha} . {\omega}^{\alpha}_{\beta} (x) + i T^{\alpha} .
{\omega}^{\alpha} (x) - i T^{\alpha} f^{\alpha\beta\gamma}
{\omega}^{\alpha\beta}_{\beta} {\omega}^{\gamma} ) .
\end{equation}

where,
\begin{displaymath}
{\omega}'^{\alpha}_{\beta} = G {\omega}^{\alpha}_{\beta} , \qquad
{\omega}'^{\alpha} = g {\omega}^{\alpha} , \qquad
W'_{\lambda} = g W_{\lambda} .
\end{displaymath}

\begin{equation}
\underline{D}_{\lambda} = {\partial}_{\lambda} - i T^{\alpha}
{\Gamma}^{\alpha}_{\lambda\beta} - i T^{\alpha} W^{\alpha}_{\lambda}
- i T^{\alpha} W^{\alpha\alpha}_{\lambda\beta} .
\end{equation}

\begin{equation}
\underline{W}^{\alpha}_{\lambda} = ( {\Gamma}^{\alpha}_{\lambda\beta} + W^{\alpha}_{\lambda} + W^{\alpha\alpha}_{\lambda\beta}) .
\end{equation}

\begin{equation}
\underline{\omega}^{\alpha} = {\omega}^{\alpha}_{\beta} + {\omega}^{\alpha} - f^{\alpha\beta\gamma} {\omega}^{\alpha\beta}_{\beta} {\omega}^{\gamma} .
\end{equation}

where,

\begin{displaymath}
{\Gamma}^{\alpha}_{\lambda\beta} = {\partial}_{\lambda} {\omega}^{\alpha}_{\beta} , \qquad W^{\alpha\alpha}_{\lambda\beta} =
f^{\alpha\beta\gamma} {\omega}^{\alpha\beta}_{\beta} W^{\gamma}_{\lambda} .
\end{displaymath}

These expressions generalize $D_{\lambda}$ , $W^{\alpha}_{\lambda}$ , ${\omega}^{\alpha}$ , in the standard Yang-Mills formalism,

\begin{equation}
\underline{W}^{\alpha}_{\lambda} \longmapsto \underline{W}^{\alpha}_{\lambda}
+ {\partial}_{\lambda} \underline{\omega}^{\alpha} - f^{\alpha\beta\gamma}
\underline{\omega}^{\beta} \underline{W}^{\gamma}_{\lambda} .
\end{equation}

\begin{equation}
\psi^{*} ( i \gamma \underline{D} - m ) \psi .
\end{equation}

Points: \\
$\partial_{\lambda} \underline{\omega}^{\alpha}$: Topological nature of the source. \\
$f^{\alpha\beta\gamma} \underline{\omega}^{\beta} \underline{W}^{\gamma}_{\lambda}$ : Topological nature of the
transformations,performed by the functional $\underline{\omega}^{\alpha}$, and, isovectors and Riemannian manifold.

\begin{equation}
W^{\alpha\alpha}_{\lambda\beta} = \frac{1}{2} g^{\alpha\mu} ( W^{\alpha}_{\lambda} g_{\mu\beta} + W^{\alpha}_{\beta} g_{\mu\lambda}
- W^{\alpha}_{\mu} g_{\lambda\beta} ).
\end{equation}

\begin{equation}
f^{\alpha\beta\gamma} \partial_{\lambda} \omega^{\alpha\beta}_{\beta} \omega^{\gamma} = \Gamma^{\alpha}_{\lambda\beta} \omega^{\alpha},
\end{equation}

\begin{equation}
g^{\alpha\mu} \partial_{\lambda} g_{\mu\beta} \omega^{\alpha} = g^{\alpha\mu} \partial_{\lambda} g_{\mu\beta} g^{\mu\beta, \alpha}
g_{\mu\beta}^{\alpha}.
\end{equation}

\section{Yang Mills Einstein Gauge Field}
From the commutator $[ \underline{D}_{\lambda} ,  \underline{D}_{\nu} ]$ , \\

\begin{eqnarray}
\begin{array}{llllllll}
F^{\alpha}_{\lambda\mu}   & + {\partial}_{\lambda} & W^{\alpha}_{\nu} & - {\partial}_{\nu} & W^{\alpha}_{\lambda} & + f^{\alpha\beta\gamma}
& W^{\beta }_{\lambda}  & W^{\gamma}_{\nu}   \\
      & + {\partial}_{\lambda} & W^{\alpha\gamma}_{\nu\delta} & - {\Gamma}^{\gamma}_{\nu\delta} & W^{\alpha}_{\lambda}
& + f^{\alpha\beta\gamma} & W^{\beta}_{\lambda} & W^{\gamma\gamma}_{\nu\delta}   \\
      & + {\Gamma}^{\alpha}_{\lambda\beta} & W^{\alpha}_{\nu} & - {\partial}_{\nu} & W^{\alpha\alpha}_{\lambda\beta}
& + f^{\alpha\beta\gamma} & W^{\beta\alpha}_{\lambda\beta}  & W^{\gamma}_{\nu}   \\
      & + {\Gamma}^{\alpha}_{\lambda\beta} & W^{\alpha\gamma}_{\nu\delta} & - {\Gamma}^{\gamma}_{\nu\delta} & W^{\alpha\alpha}_{\lambda\beta}
& + f^{\alpha\beta\gamma} & W^{\beta\alpha}_{\lambda\beta} & W^{\gamma\gamma}_{\nu\delta}      \\
      & + {\partial}_{\lambda} & {\Gamma}^{\gamma}_{\nu\delta} & - {\partial}_{\nu} & {\Gamma}^{\alpha}_{\lambda\beta}
& + f^{\alpha\beta\gamma} & {\Gamma}^{\alpha}_{\lambda\beta} & {\Gamma}^{\gamma}_{\nu\delta} .
\end{array}
\end{eqnarray}

Conformity of this structure with Yang-Mills is brought about by complexity of the transformations that $G^{\alpha}_{\beta}$ performs, in conformity
with structure of the space itself.

\begin{equation}
F^{\alpha}_{\lambda\nu} \longmapsto F^{\alpha}_{\lambda\nu} - f^{\alpha\beta\gamma} {\omega}^{\beta} F^{\gamma}_{\lambda\nu} .
\end{equation}

The linear terms in lines 3 and 4,

\begin{eqnarray}
+ {\partial}_{\lambda} g^{\alpha}_{\beta} W^{\alpha}_{\nu} + {\partial}_{\beta} g^{\alpha}_{\lambda} W^{\alpha}_{\nu} + {\partial}_{\nu} W^{\alpha\alpha}
g_{\lambda\beta} \nonumber \\
- {\partial}^{\alpha} g_{\lambda\beta} W^{\alpha}_{\nu} - {\partial}_{\nu} W^{\alpha}_{\beta} g^{\alpha}_{\lambda} - \partial_{\nu}
W^{\alpha}_{\lambda} g^{\alpha}_{\beta} .
\end{eqnarray}

\begin{displaymath}
{\nu} {\perp} \left( {\lambda} , {\beta} , {\alpha} \right) .
\end{displaymath}

\begin{eqnarray}
+ {\partial}_{\lambda} g^{\alpha}_{\beta} + {\partial}_{\beta} g^{\alpha}_{\lambda} - {\partial}^{\alpha} g_{\lambda\beta} 
( W^{\alpha}_{\nu} g^{\gamma}_{\delta} + W^{\alpha}_{\delta} g^{\gamma}_{\nu} - W^{\alpha\gamma} g_{\nu\delta} ) \nonumber \\
- {\partial}_{\nu} g^{\gamma}_{\delta} + {\partial}_{\delta} g^{\gamma}_{\nu} - {\partial}^{\gamma} g_{\nu\delta} ( W^{\alpha}_{\lambda} g^{\alpha}_{\beta}
+ W^{\alpha}_{\beta} g^{\alpha}_{\lambda} - W^{\alpha\alpha} g_{\lambda\beta} ) .
\end{eqnarray}

\begin{displaymath}
( {\nu} , {\delta} , {\gamma} ) {\perp} \left ( {\lambda} , {\beta} , {\alpha} \right) .
\end{displaymath}

Lines 2, 3, and 4, show some new properties of $\{^{\alpha}_{\beta}\}$ , as they mediate the way between the linear and the bilinear terms in
$F^{\alpha}_{\lambda\nu}$ . This can be interpreted as $\underline{\omega}^{\alpha}$ wave equation as distinct from
$\underline{W}^{\alpha}_{\lambda}$ wave equation.\\

Note: Further work on,

\begin{equation}
\underline{D}^{\alpha\beta}_{\lambda} F^{\beta}_{\lambda\nu} = g \underline{J}^{\alpha}_{\lambda} .
\end{equation}

\begin{equation}
[ \underline{D}_{\mu} , [ \underline{D}_{\lambda} , \underline{D}_{\nu} ]] + [ \underline{D}_{\nu} , [ \underline{D}_{\mu} , \underline{D}_{\lambda}
]] + [ \underline{D}_{\lambda} , [ \underline{D}_{\nu} , \underline{D}_{\mu} ]] = 0 .
\end{equation}

\begin{equation}
{\partial}_{\lambda} F^{\alpha}_{\lambda\nu} = f^{\alpha\beta\gamma} \underline{W}^{\gamma}_{\lambda} F^{\beta}_{\lambda\nu} .
\end{equation}

\end{document}


\title{Quantum Topology}
\author{Diaa A. Ahmed}
\address{e-mail vasishta@dds.nl}
\date{Cairo June 1989}

\begin{abstract}
``Quantum Topology'' deals with the general quantum theory as the theory of quantum space. On the quantum level space time and 
energy momentum forms form a connected manifold; a functional quantum space. Many problems in quantum theory and field theory 
flow from not perceiving this symmetry and the functional nature of the quantum space.\\
Both topology, groups and logic are based on the concept of sets. If properties coincide with the open sets of topology, then 
logic and topology will have the same structure. If transformations are continuous in topology, then we will have topological groups; 
we can derive fields. Therefore, quantum logic underlies the manifold and the fields and nature is based on the language of quantum logic. \\
Quantum theory and field theory based on sets and the derived topology, group and logic structures should address the question of 
computation and the mind; the quantum computer and the quantum mind. 
\pacs {03.65.-w} {11.10.-z} {11.15.-q} {04.60.-m}
\end{abstract}
\maketitle

\section{INTRODUCTION}
Minkowski said: ``{\em Henceforth space by itself and time by itself are doomed to fade away into mere shadows and only a kind of union 
of the two will preserve an independent reality}''. \\
We can take a little step and see that the natural extension to that came with the quantum theory; that the space-time form 
and the conjugate energy-momentum form are not independent realities by themselves, they recede leaving behind a functional space; 
the quantum space; then topology dominates physics.\\
The role played by $\hbar$ in quantum theory, analogous to the role of $C$ in relativity theory; it shows us more deeply the 
connectedness of the world. 
This view about the quantum is presented in the spirit of Minkowski's radical view of space and time.

\subsection{The Quantum Question}
There is a connection between quantum theory and relativity theory, take the relationship $\Delta E \Delta t \geq \hbar$ , and suppose 
we have a physical system in the vacuum, its energy will go into change, it will look as if it went into Lorentzian transformation 
$L E \longmapsto E'$ , its proper time will go into reciprocal transformation $L^{-1} T \longmapsto T'$ , such that their product; 
the action integral itself will remain invariant.  Let the system smoothly vanish $( E_{0} \longmapsto 0 )$ , now in vacuum,
fluctuations of energy will look as if it is accompanied by reciprocal fluctuations of time ( as if a gravitational field fluctuate), 
such that $\hbar$ will be our invariant. \\
Energy-momentum fluctuations in vacuum will be accompanied by reciprocal fluctuations in the metric, the whole manifold will go 
into fluctuations, energy-momentum and space-time lose heir independence and become a connected manifold, a functional-quantum space. 
We are left for $\hbar$ alone; the invariant. \\
Why $G$ ? If I may speculate about a ratio between $\hbar$ and cosmological action that makes $G$ a weak coupling on 
a cosmological scale; just a trace of the direct coupling.

\subsection{Optics Mechanics Analogy}

\begin{eqnarray}
k = \partial \varphi / \partial r , \qquad  p = \partial S / \partial r , \qquad  \hbar k = p , \nonumber \\
\omega = \partial \varphi / \partial t , \qquad  E = \partial S/ \partial t , \qquad \hbar \omega = E .
\end{eqnarray}

Purely geometric quantities $( \varphi , k , \omega )$ acquired dynamical meaning.

\subsection{Waves, Particles and Probability}
Waves that do not scatter, particles that do not localize, probabilities that interfere, the quantum sense of these 
concepts required a  generalization beyond the classical meaning of these concepts to some underlying more basic concept.

\subsection{General Relativity}
Treating energy-momentum form symmetrically with the geometric aspect of nature in the functional space will lead us from 
an incomplete geometric view to a complete topological view. \\
Functional space seems to be the way out of the singularities of space-time: Functional have no value at separate points. \\
Quantization.

\subsection{Vacuum State}
Quantum fluctuations in the vacuum endowed the vacuum with a complete picture of the dynamics. Fluctuations in energy-momentum 
in vacuum means that the vacuum is not the simple geometric aspect of nature. Simple dimensional analysis shows so; it is 
$\hbar$ space; it is $\hbar$ that generates the dynamics. We can see that, the vacuum is the functional `quantum' space. 
The pure quantum aspect of nature.

\section{QUANTUM SPACE}
T 1. Finiteness of $\hbar$: \\ 
(a)    $q \times p \neq \emptyset$ \qquad ( Means that both $q$ and $p$ are nonempty ) \\ 
(b) If $q \times p = \emptyset$ \qquad Then $q = \emptyset$ Or $p = \emptyset$ . \\ 
It is the finiteness of $\hbar$ which leads us to the functional space. In the classical limit 
$\hbar \longmapsto 0$ , (2) is valid; $p_{\mu}$ , and $q^{\mu}$ , split into separate forms; so the classical action 
can not lead to a functional space. \\ 
A 1. {\em $\hbar$ is a finite measure that generates a functional space; the quantum space}. \\ 
Our starting point is the antilinear bilinear form $( p_{\mu} , q^{\mu} )$ , a product of forms 
$\int p_{\mu} d q^{\mu}$ , that build up the functional $\hbar$ . It is obvious that this form has 
infinitely many states of representation. We can derive these states from the theory of  representation.

\section{REPRESENTATION SPACE}
T 2. Reisz Representation Theorem: \\
Each linear functional $\phi$ on a Hilbert space $H$ can be expressed:

\begin{equation}
\phi (\psi) = ( \psi_{\phi} , \psi ) ,
\end{equation}

$( \psi_{\phi} , \psi )$ is an inner product and $\psi_{\phi}$ is an element of $H^{*}$ uniquely determined by,

\begin{equation}
\psi_{\phi} = \int \phi ( \psi_{\phi})^{*} \psi_{j} d j .
\end{equation}

A 2. {\em A product of a Hilbert space $H$ with its dual space $H^{*}$ in quantum theory $< A I A >$} . \\
So, the states and the Hilbert space of the quantum theory follow directly from the functional representation, 
with the difference that, only the functional $< A I A >$ is significant.

\section{PROJECTIONS}
We can built up the space by taking the topological product $T_{1} \times T_{2}$ , of its subspaces. \\
T 3. Topological Theorem: \\
The transformations,

\begin{eqnarray}
\hat{p} : T_{1} \times T_{2} \longmapsto T_{1} , \nonumber \\
\hat{q} : T_{1} \times T_{2} \longmapsto T_{2} .
\end{eqnarray}

defined by,

\begin{eqnarray}
\hat{p} ( x_{1} , x_{2} ) = X_{1} , \nonumber \\
\hat{q} ( x_{1} , x_{2} ) = X_{2} .
\end{eqnarray}

are continuous, they are projections of the product space into its subspaces. \\
A 3. {\em In quantum space such mappings can be seen as quantum operators project the quantum space into one of its subspaces}.
We start with Fourier integral relationship,

\begin{eqnarray}
< p' I X > = \hbar^{{-}{\frac{1}{2}}} \int e^{-iq'p'/\hbar} d q' < q' I X >, \nonumber \\
< q' I X > = \hbar^{{-}{\frac{1}{2}}} \int e^{ iq'p'/\hbar} d p' < p' I X >.
\end{eqnarray}

to derive,

\begin{eqnarray}
p_{\mu} = - i \hbar \partial / \partial q^{\mu} ,\nonumber\\
q^{\mu} = + i \hbar \partial / \partial  p_{\mu} .
\end{eqnarray}

and see that; displacements:

\begin{eqnarray}
- \partial / \partial q^{\mu} \longmapsto \hat{p}_{\mu} , \nonumber \\
+ \partial / \partial p_{\mu} \longmapsto \hat{q}^{\mu} .
\end{eqnarray}

are the corresponding quantum operators; projections of the quantum space.

\begin{eqnarray}
i \hbar \hat{p}_{\mu} \longmapsto p_{\mu} , \nonumber \\
i \hbar \hat{q}^{\mu} \longmapsto q^{\mu} .
\end{eqnarray}

\section{FUNCTIONAL INTEGRALS}
We can develop the representation in terms of Dirac $\delta$-function; we take the set of basic vectors-forms and normalize,

\begin{equation}
< \xi' I \xi''> = \delta ( \xi' - \xi'') .
\end{equation}

the singular point $\xi' = \xi''$ , should be seen as the condition that only the product of a basic vector-form with its conjugate, 
$\xi' = \xi''$ , is permissible and finite ( we may call it conjugate or functional product ) only such product builds up 
a functional space, the $\delta$-function will look like an infinite diagonal matrix, represents such functional. \\
We can see that the requirement of the product being functional normalized
the space and removed certain type of infinity. \\
T 4. The $\delta$-Space:\\
$\delta$-function is a functional $( \delta , I A > )$ on the set of basic
vectors-forms. It generates a functional space. Its representative is the antilinear bilinear form $< A I A >$ , we will call it the $\delta$-space. \\
A 4. {\em Any quantum operator should be expanded in $\delta$-space}.
We have the basic equation,

\begin{equation}
( \delta , f ) = \int f (x) \delta (x) d x = f (0) .
\end{equation}

for such an expansion, instead of the usual $< X I F I X >$ of the quantum theory. It is obvious that such an expansion in functional space will lead to distributions. \\
In the system of units $\hbar = \delta = 1$ , $\hbar$-space is not different from $\delta$-space. The states themselves transform to 
quantum operators and this is the physically significant space, the process of normalization carries on automatically to the operators themselves. The functional constraint imply that only conjugate interactions are permissible in the vacuum state; that means that no such thing as interaction between indeterminate number of particles.

\section{FEYNMAN FUNCTIONAL INTEGRAL}
We have the basic equation,

\begin{equation}
( \delta , \varphi ) = \int \varphi (q) \delta (q) d q = \varphi (0) ,
\end{equation}

T 5. We can choose a convenient representation of $\delta$-function, in terms of the action integral itself,

\begin{equation}
< q' I q''> = \delta ( q' - q'') = e^{i S / \hbar} ,
\end{equation}

A 5. {\em This gives us},

\begin{equation}
\int \varphi (q) e^{ i S (q) / \hbar} d q = \varphi < O > .
\end{equation}

This allow us to expand any function of a dynamical variable in $\delta$-space in terms of the action integral. \\
( Analogous; Fourier Integral Theorem ).

\section{EXTENDED SYMMETRY}
Space-time are subject to Poincare,

\begin{equation}
x^{\mu} \longmapsto x'^{\mu} = a^{\mu} + l^{\mu}_{\nu} x^{\nu} .
\end{equation}

Extension into the functional space imply other symmetries; the symmetries of the antilinear bilinear form $( p_{\mu}$ , $q^{\mu} )$ ; 
symmetry between $p_{\mu}$ and $q^{\mu}$ . \\
T 6. Topological Theorem: \\
The graph $G$ , of a continuous transformation,

\begin{equation}
f : T_{1} \longmapsto T_{2} .
\end{equation}

is homomorphic with $T_{1}$ . The graph of the transformation is a subspace of the topological product 
$T_{1} \times T_{2}$ . We have the form $( p_{\mu} , q^{\mu} )$ , $p_{\mu}$'s , transform as Dirac spinors 
$p_{\mu} = i \gamma^{\mu} \partial_{\mu}$ ,

\begin{equation}
\psi' (x) = exp ( - {\frac{i}{4}} \sigma^{\mu\nu} \varepsilon_{\mu\nu} ) \psi(x).
\end{equation}

transformations of the form $( p_{\mu} , q^{\mu} )$ ,

\begin{equation}
p_{\mu} q^{\mu} \longmapsto p'_{\mu} q'^{\mu} .
\end{equation}

A 6. {\em Will mix four Minkowski coordinates and four Dirac spinor displacements, the resulting group will be homomorphic with 
Poincare group}. People extensively explored such symmetry under the name ``Supersymmetry''.\\
$C$ showed us the connectedness of space and time. $\hbar$ showed us the connectedness of $p_{\mu}$ and $q^{\mu}$ in the quantum space, 
it shows us the continuum that lies behind the discrete nature, reality itself is the quantum space, simple and beautiful.